\documentclass{PoS}
\usepackage{amsmath,fontenc,calrsfs}
\usepackage{amsfonts,amssymb,mathtools,slashed}

\def\beq{\begin{equation}}
\def\eeq{\end{equation}}
\def\bea{\begin{eqnarray}}
\def\eea{\end{eqnarray}}
\def\nn{\nonumber}

\def\chic1{\chi_{c1}}

\newcommand{\vep}{\varepsilon}

\def \Im{\text{Im}\,}

\def\Xint#1{\mathchoice
   {\XXint\displaystyle\textstyle{#1}}%
   {\XXint\textstyle\scriptstyle{#1}}%
   {\XXint\scriptstyle\scriptscriptstyle{#1}}%
   {\XXint\scriptscriptstyle\scriptscriptstyle{#1}}%
   \!\int}
\def\XXint#1#2#3{{\setbox0=\hbox{$#1{#2#3}{\int}$}
    \vcenter{\hbox{$#2#3$}}\kern-.5\wd0}}

\def\dashint{\Xint-}

\newcommand{\vx}{\mathbf{x}}

\newcommand{\vq}{\mathbf{q}}

\title{Pole structure and compositeness}

\ShortTitle{Pole structure and compositeness}

\author{\speaker{J.~A.~Oller}\thanks{This work is supported in part by the MINECO (Spain) and EU grant FPA2016-77313-P.}\\
        Author Departamento de F\'\i sica,
Universidad de Murcia, E-30071 Murcia, Spain\\
        E-mail: \email{oller@um.es}}

\abstract{We present in this talk a series of new results on the nature of a bound state or
resonance based on the calculation 
  of the expectation values of the number operators of the free particles in the
  state of interest. In this way,  a new universal criterion for the elementariness
  of a bound state emerges. In the case of large particle wavelengths 
 compared to the range of their interaction,  a new
  closed formula for the compositeness of a bound state in a two-particle continuum is obtained.
  The extension of these results to resonances with respect to the open channels
  can be given  by making use in addition of suitable
  phase-factor transformations as also reviewed  here.  We end  with a discussion on the $X(3872)$
as possible double- or triple-pole virtual state, which would be the first case in particle phenomenology.
}

\FullConference{XVII International Conference on Hadron Spectroscopy and Structure - Hadron2017\\
		25-29 September, 2017\\
		University of Salamanca, Salamanca, Spain}

\begin{document}

\section{Introduction}
\label{ref.170929.0}

 We have given in Ref.~\cite{oller.171114.1} a new perspective to the problem of the compositeness/elementariness of
 a bound state or a resonance by considering the expectation values in the state of the number operators
 of the free particle species.
 The essential problem is to discern whether this bound state is elementary or composite;
 because of the possible impact of underlying degrees of freedom this is not always a question of an easy answer.
 In many applications within
effective field theory (EFT) the bare elementary discrete states are typically integrated out and do not
appear explicitly in the Lagrangian of the theory. Nonetheless, one can still generate
bound states after complementing the perturbative calculations in the corresponding EFT with
non-perturbative techniques \cite{Weinberg.170929.1,oller.211116.5}.
 In particular, a near-threshold bare elementary discrete state can be mimicked by including a Castillejo-Dalitz-Dyson
pole (CDD)  \cite{Castillejo.171020.1}, which straightforward manifestation is an ``innocent'' zero in a partial wave amplitude 
 of the free continuum states.
 Explicit examples are
worked out in Refs.~\cite{oller.211116.5,kang.170930.1}.
It is also the case that the Hamiltonian
might be expressed in terms of degrees of freedom that are not asymptotically free, as it
occurs in Quantum Chromodynamics in ordinary conditions.
 In all these cases it is not a priori clear whether an
eigenstate of the full Hamiltonian is composite or elementary with regards to the asymptotic states in the continuum.

Let us discuss first the case of a bound state within non-relativistic quantum mechanics (NRQM) \cite{ref.170928.1,ref.170928.2} and let us split the full Hamiltonian $H$ in an
unperturbed free-particle part $H_0$ and an interaction $V$,
$  H=H_0+V$.  The spectrum of the full Hamiltonian consists of the continuum states
and it might contain also discrete bound states $|\psi_{n}\rangle$, and similarly for $H_0$ (in which
case the discrete states are bare elementary ones):
\begin{align}
  \label{170928.2}
  \begin{array}{lllll}
    H|\psi_\alpha\rangle =E_\alpha|\psi_\alpha\rangle & & &&  H_0|\varphi_\alpha\rangle =E_{\alpha}|\varphi_\alpha\rangle \\
 H|\psi_{n}\rangle =E_{n}|\psi_{n}\rangle & & &&  H_0|\varphi_n\rangle =E_{n}|\varphi_n\rangle
  \end{array}
\end{align}
Both $H$ and $H_0$ share the same spectrum \cite{ref.170928.3,ref.170928.4}.
 One should not confuse the masses of the  eigenstates in $H_0$
 with bare masses present in a Lagrangian \cite{ref.170928.3};
 the difference if any must be included in $V$.

 Given a bound state $|\psi_B\rangle$ of $H$  
we express it in terms of the eigenstates
 of $H_0$, which fulfill a completeness relation, as
 \begin{align}
   \label{170929.1b}
   |\psi_B\rangle&=
   \int d\alpha \langle \varphi_\alpha|\psi_B\rangle |\varphi_\alpha\rangle
   +\sum_n \langle \varphi_n|\psi_B\rangle |\varphi_n\rangle~,\\
   \langle \psi_B|\psi_B\rangle&=1=\int d\alpha |\langle \varphi_\alpha|\psi_B\rangle|^2
   +\sum_n |\langle \varphi_n|\psi_B\rangle|^2=Z+X~,\nn
 \end{align}
 where
 \begin{align}
   \label{170929.4}
   X&=\int d\alpha |\langle \varphi_\alpha|\psi_B\rangle|^2~,\\
   \label{170929.4b}
   Z&=\sum_n |\langle \varphi_n|\psi_B\rangle|^2~.
 \end{align}
 These quantities are usually called compositeness ($X$) and
 elementariness ($Z$).

\section{A different perspective on the compositeness of a bound state}
\label{ref.170929.1}

We first start with the non-relativistic (NR) case and later we move to the relativistic one. 
For definiteness, let us take two particle species $A$ and $B$ whose free-particle annihilation/creation
operators are denoted by  $a_\alpha/a_\alpha^\dagger$ and $b_\beta/b_\beta^\dagger$, respectively.
The decomposition of the bound state in eigenstates of $H_0$, Eq.~\eqref{170929.1b}, reads now
 $   |\psi_B\rangle=
   \int d\gamma \langle AB_\gamma|\psi_B\rangle |AB_\gamma\rangle
   +\sum_n \langle \varphi_n|\psi_B\rangle |\varphi_n\rangle$.
For a given particle species $A$ its number operator is denoted by $N_D^A$ and given by
$N_D^A=\int d\alpha \,a_\alpha^\dagger a_\alpha.$ 
 Here the subscript $D$ refers to the Dirac or interaction image. Notice that since $N_D$ and $H_0$
obviously commute then $N_D^A(t)=e^{i H_0 t} N_D^A(0) e^{-i H_0 t}=N_D$.
 Based on the number operators of $A$ and $B$  we define the compositeness $X$ of the bound state $|\psi_B\rangle$ as
\begin{align}
  \label{170929.11}
  X&=\frac{1}{2}\langle \psi_B|N_D^A+N_D^B|\psi_B\rangle~.
\end{align}
That is, $X$ is the expectation value of the number operator of the
 free-particle constituents in the eigenstate $|\psi_B\rangle$ of $H$ divided by
their nominal  number, which in this case is 2.
 The new definition of $X$ is equivalent to the original one of Eq.~\eqref{170929.4} because 
 $(N_D^A+N_D^B)|AB_\gamma\rangle=2 |AB_\gamma\rangle$ and the annihilation operators $a_\alpha$ and $b_\beta$
 destroy the bare elementary discrete states. 
 It then follows that  $X$, as defined in 
 Eq.~\eqref{170929.11}, reads  $  X=\int d\gamma |\langle AB_\gamma|\psi_B\rangle|^2~,$
 as in Eq.~\eqref{170929.4}. In general, if we are applying NRQM to a bound state $|\psi_B\rangle$ of 
$n$ particles corresponding to $m$ particle species $A_1$, $\ldots$, $A_m$,
the compositeness is defined by a straightforward generalization of
the two-body case of Eq.~\eqref{170929.11} as 
\begin{align}
  \label{170929.13}
  X&=\frac{1}{n}\langle \psi_B|\sum_{i=1}^m N_D^{A_i}|\psi_B\rangle~.
\end{align}


The evaluation of the expectation values for the number operators is
amenable to a direct computation  within NR Quantum Field Theory (QFT).
  At time $t$ the states $|\varphi(t)\rangle$ evolves in the Dirac picture by the time translation
  $|\varphi(t)\rangle=U_D(t,0)|\psi\rangle$,  $U_D(t_2,t_1)=e^{iH_0 t}e^{-iHt}e^{-iH_0t}$.
In particular, the bound state $|\psi_B\rangle=|\varphi_B(0)\rangle$  can be expressed  by the
time evolution from the asymptotic  bare elementary discrete state $|\varphi_B\rangle$ as
\begin{align}
  \label{170929.15}
  |\psi_B\rangle&=U_D(0,\pm \infty)|\varphi_B\rangle~.
  \end{align}
This allows us to write $X$ in a time-ordered way by
introducing an extra time evolution from 0 to $t$ as \cite{oller.171114.1}
\begin{align}
  \label{170930.4}
    X&=\frac{1}{n}\lim_{T\to +\infty}\frac{1}{T}\int_{-T/2}^{+T/2} dt \langle \varphi_B|U_D(+\infty,t)
    N_D(t)U_D(t,-\infty)|\varphi_B\rangle~,
\end{align}
  with $N_D=\sum_{i=1}^m N_D^{A_i}$. The factor $1/T$ 
 cancels in the limit $T\to +\infty$ with the Dirac delta function of total energy
conservation. It might be advantageous to express the number operator in terms of NR fields in
Eq.~\eqref{170930.4}, e.g. in order to apply Feynman diagrams for its calculation.
 For a generic scalar particle species $A_i$ of physical mass $m_{A_i}$
we have the free field $\psi_{A_i}(x)$, $x=(t,\vx)$, and we can write
 $N_D=\sum_i \int d^3\vx\, \psi^\dagger_{A_i}(x)\psi_{A_i}(x)~.$
 Inserting this expression into Eq.~\eqref{170930.4} it reads
\begin{align}
  \label{171110.3}
 X &=\frac{1}{n}\lim_{T\to +\infty}\frac{1}{T}\int d^4x 
\langle \varphi_B| P\left[ e^{-i\int_{-\infty}^{+\infty}dt' V_D(t')}
 \sum_i \psi_{A_i}^\dagger(x)\psi_{A_i}(x)
 \right] |\varphi_B\rangle~.
  \end{align}
Here we denote the time-ordered product by $P$ and $V_D(t)$ is the interaction in the
Dirac picture.\footnote{In Eq.~\eqref{171110.3} only the connected diagrams should be considered
  \cite{Low.171110.1}.} The extension to particles with other spin is straightforward.

\begin{figure}
\begin{center}
\includegraphics[width=0.6\textwidth]{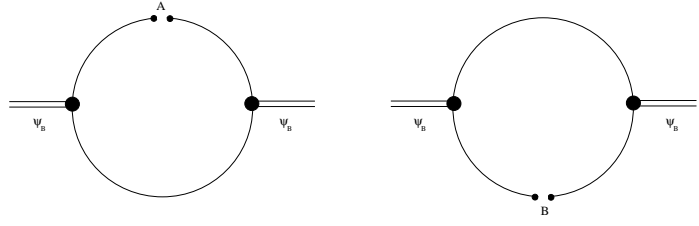}
\caption{Feynman diagrams for the calculation of $X$ within NR QFT
  for the two-particle case. The insertion of the number operators for
the particles $A$ and $B$ is indicated by the double dot.}
\label{fig.170930.1}
\end{center}
\end{figure}

For two particles of types $A$ and $B$,
the evaluation of $X$ corresponds to  the calculation of the diagrams
in Fig.~\ref{fig.170930.1}. In the $\ell S$ basis (with $\ell$ the orbital angular momentum and
$S$ the total spin) one has, $  X=\sum_{\ell,S} X_{\ell S}$, 
\begin{align}
  \label{170930.6}
  X_{\ell S}&=\frac{1}{2\pi^2}\int_0^\infty dk k^2 \frac{g_{\ell S}^2(k^2)}{(k^2/2\mu-E_B)^2}~.
\end{align}
In this equation, $\mu$ is the reduced mass of particles $A$ and $B$, and 
$g_{\ell S}^2(k^2)$ is the coupling  squared of the
bound state, $\langle AB_{\ell S}|V|\psi_B\rangle^2$.
The coupling can be calculated from the knowledge of the residue at $E=E_B$ of the
 the off-shell  $T$ matrix, $T(E)(k',k)$, and it satisfies then the following integral equation
(a matrix notation should be employed if appropriate) \cite{oller.171114.1} 
\begin{align}
  \label{170930.8}
  g(k)&=\frac{1}{2\pi^2}\int_0^\infty dk'\,{k'}^2V(k,k')\frac{1}{E_B-{k'}^2/2\mu}g(k')~.
\end{align}
 From Eq.~\eqref{170930.8} and the fact that $V(-k,k')=(-1)^\ell V(k,k')$ (parity conservation)
 one concludes that the coupling squared only depends on $k^2$. 
  The global normalization factor in Eq.~\eqref{170930.8}
  is fixed by the requirement that $g(k)$
  matches the residue of the $T$ matrix at the pole position.
  From Eq.~\eqref{170930.8} it turns out that
 $g(k)$ is analytic in the $k$-complex plane without cuts \cite{oller.171114.1}. 
 As a result, a non-constant $g(k)$ is not bounded for $k\to \infty$
 because of the Liouville's theorem in complex analysis.

 It is worth stressing that for a given total Hamiltonian $H$ the compositeness $X$ is 
 an  observable in the sense that it is invariant under unitary transformations and field
 reparametrizations. This is clear from the decomposition of the bound state in terms of free-particle
 states. 
 The absence of tad-pole like contributions within an appropriate regularization procedure
 in NR QFT drives to Eq.~\eqref{170930.6} as the final expression without any possible counterterm
 contributions \cite{oller.171114.1}. 
 Thus, $X$ is a fully derived quantity from the knowledge of the (half) off-shell $T$ matrix (which
 contains all the spectroscopical information of the corresponding quantum system).

We derive now a new closed expression for the compositeness $X$ in the case in which the wavelengths of the
 two scattering particles are large compared with the range of their interaction.
In this case, $V(k',k)$ is a polynomial in its arguments  and
$|g(k)|$ grows only polinomically for $k\to \infty$.
A simple way to deal with these power-like divergences is to regularize the potential as
 $  V(k',k)\to V(k',k) e^{i\epsilon (k+k')}$, 
with $\epsilon \to 0^+$. 
It is clear from Eq.~\eqref{170930.8} that this also implies that $g(k) \to g(k) e^{i\epsilon k}$~.
Taking advantage of the fact that the integrand in Eq.~\eqref{170930.6}
is an even function of $k$ we symmetrize it, extend the integration along the real axis
from $-\infty$ to $+\infty$, and
close the integration contour  along the
upper half plane of the $k$-complex plane with a semicircle of infinite radius.
 The calculation is straightforward by applying the Cauchy's integration theorem
with the result
\begin{align}
  X&=g^2(\varkappa^2)\frac{\mu^2}{2\pi\bar{\gamma}}
  +\frac{\mu^2}{2\pi}\left.\frac{\partial g^2(-\gamma^2)}{\partial \gamma}\right|_{\gamma=\bar{\gamma}}~.
  \label{170930.11}
\end{align}
where $\gamma=-ik$ and $\bar{\gamma}=-i\varkappa$, and $\varkappa=\sqrt{-2\mu E_B}$. 
    The first term on the right hand side of this equation is a well-known
    contribution, although obtained with less generality
 \cite{ref.170928.2,hyodo.170930.1}.
    It is model independent because 
    it is fixed once the pole position and the residue of the on-shell $T$ matrix at the pole position
  are known. 
  The second term of Eq.~\eqref{170930.11} is an extra contribution, which
  cannot be fixed directly from the knowledge of the on-shell $T$ matrix
    and  depends on the interaction $V(k',k)$. 
 Let us apply  Eq.~\eqref{170930.11} to  some 
 energy-independent potentials for which   $X=1$. The first one is a pure $S$-wave
 potential 
 $V(k',k)=\left[v_0+v_2 (k^2+{k'}^2)\right]e^{i\epsilon(k+k')}$, 
 where $v_0$ and $v_2$ are constants.
 The scattering problem can be easily solved \cite{oller.171114.1} and  the two terms that sum up $X=1$ in Eq.~\eqref{170930.11} are, in order,
 \begin{align}
   \label{171107.7}
1&=\frac{1-2\bar{\gamma}^2 v_2/v_0}{1-6\bar{\gamma}^2 v_2/v_0}
 -\frac{4\bar{\gamma}^2 v_2/v_0}{1-6\bar{\gamma}^2 v_2/v_0}~.
\end{align}
 In this case, since $g^2(k^2)$ is not zero for $k=0$, the last contribution in Eq.~\eqref{171107.7} is
 suppressed by a factor $\bar{\gamma}^2 |v_2/v_0|\sim \bar{\gamma}R$.
 The last step is based on the relation between $v_0$ and $v_2$ with the effective
 range parameters \cite{oller.171114.1}.\footnote{The
   only exception to this rule might happen when there is  a shallow
   CDD pole close enough to threshold \cite{kang.170930.1}.
 In such a case  the $s$-channel exchange of the bare elementary state 
 cannot be traded by a perturbative expansion in momenta through contact interactions employing the NR equations
 of motion \cite{oller.171112.1}. The compositeness $X$ is then smaller than 1.
  \label{footnote.171107.1}}
 The Weinberg's formula for  the
 compositeness of an $S$-wave shallow bound state state  is recovered
 \cite{ref.170928.2} once the subleading contribution is neglected.
  As a specific example for which $g^2(k^2)=0$ at $k^2=0$, 
  let us take a projected potential with orbital angular momentum $\ell$,
 $V(k',k)= v_\ell {k'}^\ell k^\ell e^{i\epsilon (k+k')}$,
 with $v_\ell$ being a constant.
  The two terms  in Eq.~\eqref{170930.11} are now, in order,
\begin{align}
  \label{171107.12}
  1&=\frac{1}{2\ell+1}+\frac{2\ell}{2\ell+1}~.
\end{align}
Both contributions count on the same footing, but as $\ell$ increases the 2nd one
 becomes indeed  dominant. If the range of
 the interactions between the non-relativistic particles is explicitly resolved,
 the formula for $X$ of Eq.~\eqref{170930.11},
derived in the limit of large wavelengths compared
to $R$, is not valid in general. Of course, Eq.~\eqref{170930.6} is still applicable.
 The reason is that closing the integration contour
to end with Eq.~\eqref{170930.11} is not always possible \cite{oller.171114.1}.


 Up to the best of our knowledge there is no a general criterion for a relativistic bound state to be
 qualified as elementary. In the relativistic case one generally relies on the study of the wave function
 renormalization $Z$ \cite{amado.170930.1,salam.170930.1,lurie.170930.1}.
 The straightforward extrapolation of the definition of $X$ in
 Eq.~\eqref{170929.13} cannot be given because contributions of multi-particle eigenstates of $H_0$.
 In this way, Eq.~\eqref{170929.1b} now generalizes to
 \begin{align}
  \label{170930.12}
  |\psi_B\rangle &= \int d\gamma C_\gamma |AB_\gamma\rangle
       +\int d\eta D_\eta |AAB_\eta\rangle+\int d\mu \,\delta_\mu |ABB_\mu\rangle+\ldots\\
       &+\int d\eta_\nu F_\nu|CD_\nu\rangle+\ldots
       +\sum_n C_n|\varphi_n\rangle+\sum_n \int d\alpha C_{n\alpha} |A_\alpha\varphi_n\rangle
       +\ldots+\sum_{n,m} C_{nm} |\varphi_n\varphi_m\rangle+\ldots \nn
 \end{align}
 with quite an obvious notation.
 Nonetheless, we can still take advantage of the use of the number operators which are defined
 in the relativistic case as in NR QFT. E.g. the average number of
 asymptotic particles of type $A$ in $|\psi_B\rangle$ as given by the decomposition in
 Eq.~\eqref{170930.12} is $\langle \psi_B|N_D^A|\psi_B\rangle=\int d\gamma |C_\gamma|^2 
   +2\int d\eta |D_\eta|^2 +    \int d\mu |\delta_\mu|^2+\ldots+\sum_n \int d\alpha
   |C_{n\alpha}|^2+\ldots $
 In this way we can deduce the following universal criterion for a bound state
 to be considered as elementary, applicable both in the relativistic and NR cases:
 \begin{align}
   \label{170930.14}
   \langle \psi_B|N_D^A|\psi_B\rangle&=0~~,~~\forall A~;~~
  \langle \psi_B|N_D^E|\psi_B\rangle = 1~.
 \end{align}
 where $N_D^E$ is the sum of the number operators for the bare elementary discrete states ($N_D^n$).
 The last condition in Eq.~\eqref{170930.14} avoids
 the possibility that $|\psi_B\rangle$ had components  of states made by several bare
elementary discrete states. If for  particle species $A$ one has that  
 $ \langle \psi_B|N_D^A|\psi_B\rangle=x_A$, 
with $x_A\geq m$ and $m\geq 0$ a natural number, then  the
 free-particle states containing $m$ or more  asymptotic particles of type $A$ are relevant
 in the bound state $|\psi_B\rangle$.

\begin{figure}
\begin{center}
\includegraphics[width=0.25\textwidth]{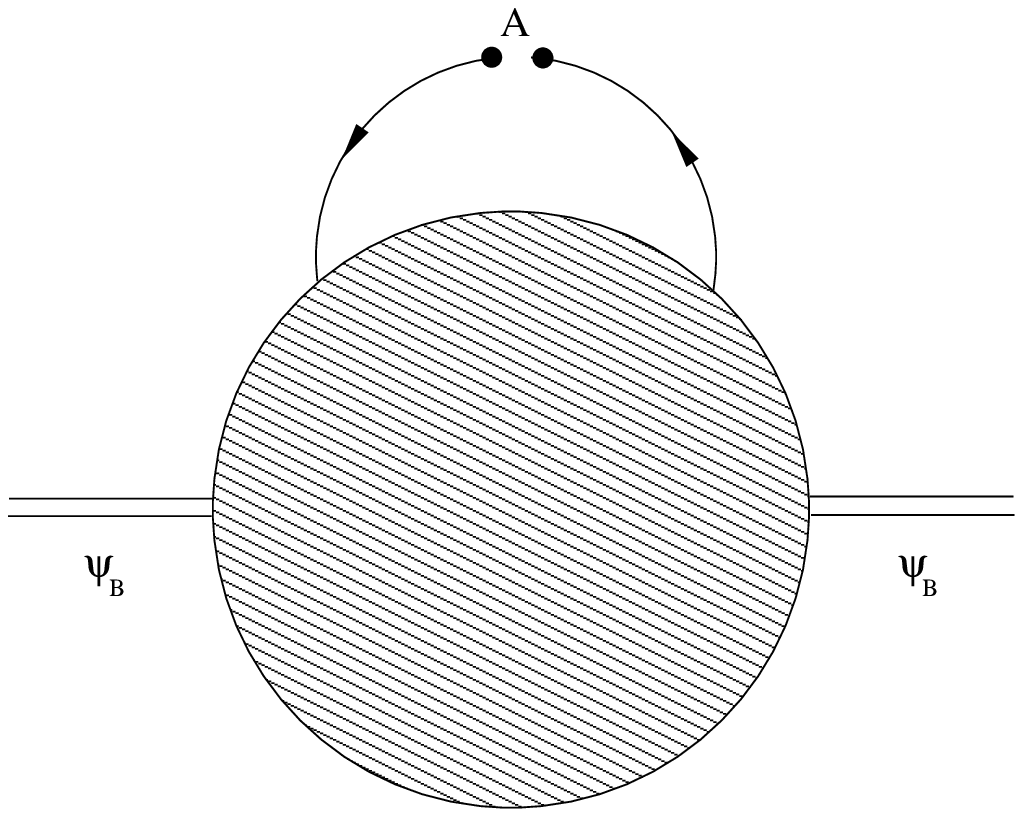}
\caption{Schematic representation of the Feynman diagrams for the calculation of
  $\langle \psi_B|N_D^A|\psi_B\rangle$ in QFT.}
\label{fig.171110.1}
\end{center}
\end{figure}

The expression for the expectation value $\langle \psi_B|N_D^A|\psi_B\rangle $ in
relativistic QFT can be obtained following similar  steps as for NR QFT \cite{oller.171114.1}, cf.
Eq.~\eqref{170930.4}.
 One can also express the number operator $N_D^A$ in terms of free fields, analogously as done in
the non-relativistic case. Let us a consider a scalar particle $A$, we can write
 $  N_D^A(t)=-2i\int d^3 \vx \dot{\psi}^{(-)}(x)\psi^{(+)}(x)$, 
\begin{align} 
  \label{171110.4}
  \psi^{(+)}(x)&=\int\frac{d^3\vq}{(2\pi)^3}a(\vq)e^{-i\tilde{q} x}~,\nn\\
\langle \psi_B|N_D^A|\psi_B\rangle   &=-2i\lim_{T\to +\infty}\frac{1}{T}\int d^4x 
\langle \varphi_B| P\left[ e^{-i\int d^4 x' {\cal H}_D(x')}
  \dot{\psi}^{(-)}(x)\psi^{(+)}(x) \right] |\varphi_B\rangle~.
\end{align}
where $\psi^{(-)}={\psi^{(+)}}^\dagger$ and the interaction have been written in terms of an interaction-Hamiltonian density
${\cal H}_D(x)$ in the Dirac picture. The set of Feynman diagrams involved
can be schematically represented as in Fig.~\ref{fig.171110.1}, where the shaded circle
represents any set of connected vertices without any insertion of the number operator which is
indicated by the double dot.
 However, in order to apply Feynman rules to the calculation of Eq.~\eqref{171110.4}
one has to take into account that any of the two internal lines in Fig.~\ref{fig.171110.1}
ending in the double dots is not a standard Feynman propagator but
$ \frac{i}{2E_k(k^0-E_k+i\vep)}$.
If one can conclude that only two-body channels  dominate
 then one expects that the more important Feynman diagrams are those of Fig.~\ref{fig.170930.1}. 
 It is worth stating that for a given total Hamiltonian $H$ the expectation values $N_D^A$ are
 observable in the sense that they are invariant under unitary transformations and field
reparametrizations, analogously as in NR QFT, see  Ref.~\cite{oller.171114.1} for further details

 \section{Resonances}
 \label{sec.170930.4}

 In this section we discuss the generalization of many of the results given for bound states
  to the case of resonances. The latter correspond to poles in an unphysical RS that can be
 reached by the analytical extrapolation in energy of the $T$ matrix.
 An approximate way to afford the problem of the evaluation of $Z$ for an unstable particle in the non-relativistic case
 near a two-body threshold was considered in Ref.~\cite{markusin.170930.2}.
  These results have also a clear connection with the
 counting pole rule of Morgan \cite{morgan.170930.1} and with the  presence of near Castillejo-Dalitz-Dyson
 poles \cite{kang.170930.1}.

\begin{figure}
\begin{center}
\includegraphics[width=0.3\textwidth]{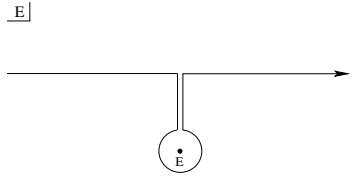}
\caption{Deformation of the integration contour along the physical energy  needed to reach the resonance
pole at $E_R=M_R-i\Gamma/2$.}
\label{fig.171001.1}
\end{center}
\end{figure}

Our definition of compositeness $X$ is 
 based on  Eq.~\eqref{170929.13} in NRQM.
 We derive our results for resonance states by the analytical continuation in energy 
 of $\langle \psi^-_\alpha|N_D|\psi^+_\alpha\rangle$ extrapolated  to the resonance pole position, with
 $|\psi_\alpha^{\pm}\rangle $ an in/out state, respectively.
 The previous matrix element has a double pole at the resonance pole whose residue divided by the coupling
 squared is the expectation value of the operator $N_D$ in the resonance state  \cite{albaladejo.171001.1}.
 This limit process can be avoided if we use an analogous formalism to
 that explained in Sec.~\ref{ref.170929.1},
 but now for resonance states. We express the in/out resonance state $|\psi^\pm_R\rangle$ by evolving the
 bare one $|\varphi_R\rangle$ from asymptotic times, $  |\psi^\pm_R\rangle =U_D(0,\mp\infty)|\varphi_R\rangle$.
 Thus an equation analogous to Eq.~\ref{170930.4} results  \cite{oller.171114.1}.
 We can also  re-express $N_D$  in
terms of bilinears of the NR fields $\psi_i(x)$.
Therefore, for scalar particles $A_i$ we have $N_D=\sum_i \int d^3\vx \psi_{A_i}^\dagger(x)\psi_{A_i}(x)$
 and we end with an expression  for $X$ analogous to  Eq.~\eqref{171110.3}  \cite{oller.171114.1}. 

In the case of two-particle asymptotic states the calculation of $X$ can be done by evaluating the
Feynman diagrams of Fig.~\ref{fig.170930.1}. Performing the corresponding partial-wave decomposition
 one has the following expression for $X_{\ell S}$
\begin{align}
  \label{171001.7}
  X_{\ell S}&=\frac{1}{2\pi^2}\int_0^\infty dk k^2\frac{g_{\ell S}^2(k^2)}{(k^2/2\mu-E_R)^2}
  +\frac{i\mu^2}{\pi\varkappa}\left.\frac{\partial kg_{\ell S}^2(k^2)}{\partial k}\right|_{k=\varkappa}~.
  \end{align}
Compared with Eq.~\eqref{170930.6} there is an extra term due to the deformation of the analytical contour
for integration, as shown in Fig.~\ref{fig.171001.1}.
 Because of the same reason, the homogeneous integral
 equation satisfied by $g(k)$ has an extra term compared to Eq.~\eqref{170930.8}, as derived in Ref.~\cite{oller.171114.1}.
    Although Eq.~\eqref{171001.7} is not explicitly real and positive,
 it is shown in Ref.~\cite{ref.171001.2} that for
 an energy-independent regular potential $X=1$, which implies that a resonance in ordinary
 QM is a composite state. The case of a separable potential 
 $V(k,k';E)=v(E) f(k^2)f({k'}^2)$ is explicitly worked out in Ref.~\cite{oller.171114.1}. 
 There it is shown that if $\partial v(E)/\partial E=0$ then 
 $X=1$, but for $\partial v(E)/\partial E\neq 0$ the compositeness $X$ is in general
a complex number.

We can also derive a new closed formula for $X$ in the case of
large wavelengths of the two scattering particles compared with the range $R$ of their
interactions.
 Since the resonance pole lies in the 2nd RS, with ${\rm Im}k<0$,
 now we regularize the potential as
$   V(k',k)\to V(k',k) e^{-i\epsilon (k+k')}$. It is straightforward to obtain
\begin{align}
  \label{171001.9}
  X_{\ell S}&=g^2(\varkappa^2)\frac{i\mu^2}{2\pi\varkappa}
  +\frac{i\mu^2\varkappa}{\pi}\left.\frac{\partial g^2(k^2)}{\partial k^2}\right|_{k=\varkappa}~.
\end{align}
The first term is already well-known while the latter is a new contribution. 
 The result for $X$ in Eq.~\eqref{171001.9} can also be expressed as
\begin{align}
  \label{171001.10}
  X_{\ell S}&=\frac{2\mu^2}{\pi^2}\int_0^\infty
 dk^2 \sqrt[{\rm II}]{k^2+i\vep}\frac{g_{\ell S}^2(k^2)}{(k^2-\varkappa^2)^2}~,
\end{align}
where $\sqrt[II]{z}$ is $\sqrt{z}$ in the 2nd RS with ${\rm arg}z\in [2\pi,4\pi[$.
    In this equation it is clear the appearance of the square of the resonance wave function
 $g_{\ell S}^2(k^2)/(k^2-\varkappa^2)^2$ 
     as it corresponds to a Gamow state \cite{ref.171001.2}.
    For the relativistic case we can evaluate the matrix elements of the operator numbers
$N_D^A$ between resonances states, and end with a similar expression to Eq.~\eqref{170930.4} \cite{oller.171114.1}.
We can also express the number operators $N_D^A(t)$ as bilinear of relativistic
fields, cf. Eq.~\eqref{171110.4}. The set of Feynman diagrams is represented in Fig.~\ref{fig.171110.1},
with the obvious replacement
of $|\psi_B\rangle$ by $|\psi_R^{\pm}\rangle$ to the right and left, respectively.
 We have to keep in  mind the meaning of the internal lines
joining the field bilinear associated to the number operator, as discussed in Sec.~\ref{ref.170929.1}.
 Since the annihilation operators in the number operators $N_D^A$
 kill the bare elementary  discrete states we can state that a necessary
 condition in relativistic (NR) QFT  for a resonance being elementary is that 
\begin{align}
  \label{171002.1}
  \langle \psi_R^-|N_D^A|\psi_R^+\rangle&=0 ~~,~~\forall A~.
\end{align}
It cannot be qualified as sufficient condition as well because given a decomposition of a resonance state
 as in Eq.~\eqref{170930.12}, one should not expect
 to have the sum of the modules squared of the coefficients in the linear decomposition
 but rather  coefficients squared \cite{oller.171114.1}.


 Let us  consider   energy-dependent transformations in the partial-wave projected in/out states.
 These are driven   by a function $\eta_i(E)$ for the $i_{\rm th}$ partial wave, which at least
 has a unitarity cut and satisfy the Schwarz reflection principle $\eta_i(E\pm i\vep)=\eta_i(E\mp i\vep)^*$. Namely,
 \begin{align}
   \label{171001.16b}
 |\psi_\alpha^+\rangle&\to e^{\eta_i(E_\alpha+i\vep)}|\psi_\alpha^+\rangle~,\\
 \langle \psi_\alpha^-|&\to \langle \psi_\alpha^-| e^{\eta_i(E_\alpha-i\vep)^*}=\langle \psi_\alpha^-|e^{\eta_i(E_\alpha+i\vep)}~.
 \end{align}
 When performing the analytical extrapolation to the 2nd RS to reach the resonance pole
 at $E_R$ one has to cross the unitarity cut and end with the value
 $\eta^{{\rm II}}(M_R-i\Gamma/2)$. As a result  the couplings change as $g_i^2(k^2)\to g_i^2(k^2) e^{2\eta_i^{{\rm II}}(E_R)}$.
 For the case of a narrow resonance we can write a plausible dispersion relation for
 the smooth function $\eta_i(E)$ around the resonance region as
 \begin{align}
   \label{171001.19}
   \eta_i(E)&=\frac{1}{\pi}\int dE'\frac{{\rm Im}\eta_i(E')}{E'-M_R-i\Gamma/2}
   \approx \frac{1}{\pi}  \dashint dE'\frac{{\rm Im}\eta_i(E')}{E'-M_R}
   +i\,{\Im}\eta_i(M_R)~.
 \end{align}
 Since ${\rm Im}\eta_i(E')$ is uniform  around the narrow-resonance mass, 
 its Cauchy principal value around  should be very small
 and  Eq.~\eqref{171001.19} is quite a pure imaginary number. The couplings then change
 as $g_i^2(k^2)\to g_i^2(k^2) e^{2i{\rm Im}\eta_i^{{\rm II}}(E_R)}$.
 This derivation also shows that for a finite width resonance is not so clear that $\eta_i(E)$
 is just a purely imaginary number. However,  if
 the resonance is manifest on the physical real energy axis, for some physical process, the modules of its residues
 can be interpreted as physical
 couplings and the corrections on them (if any) would be relatively small.
 In this way,  we can then properly choose the phase factor $\eta_i^{{\rm II}}(E_R)$ of the
 coupling to a partial wave so that its compositeness is $|X_{\ell S}|$.
 As a result,  $|X|\ll 1$ is
 the criterion for the elementariness of a NR resonance (qualified in the above sense) with respect
 to the explicit open channels.
  In the relativistic case the situation is a priori
 less clear since one cannot exclude contributions from closed channels containing particles of type $A$. Therefore,
 the change of phase in the couplings of only the open channels is not enough in general  to end with real and positive
 expectation values of $N_D^A$.
 Nonetheless, in practical applications within models that incorporate only a few coupled channels
 one could still apply these changes of phase in the couplings for the open channels and give
 physically reasonable results.
  The phase transformation here discussed is closely connected to the
  transformations in the partial-wave projected $S$-matrix introduced in Ref.~\cite{guo.170930.1}.

     \begin{figure}[ht]
\begin{center}
 \includegraphics[angle=0.0,width=.3\textwidth]{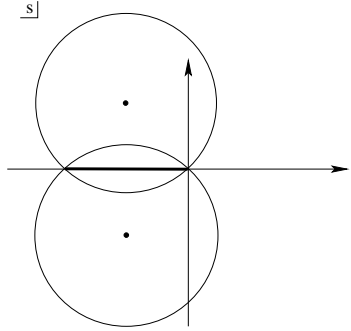}
\end{center}
\caption{Double-pole virtual state case in the $s$-complex plane .
  The pole positions are indicated by the full circles and
 the convergence radius for the Laurent series around each of them is the radius of every circle.
      \label{fig.171124.1}}
     \end{figure}
     
If a resonance has a finite width and its mass is smaller than threshold, which can happen for an $S$-wave \cite{pelaez.171125.1},
then the resonance signal is not directly manifest apart from a cusp effect.
In such case the condition above to apply the phase transformations does not apply.
If the width is small we are near the realization of a double-pole virtual state, as represented
in Fig.~\ref{fig.171124.1}, where the poles lie in the 2nd RS and the circles correspond
to the radius of convergence of the Laurent series around the poles.
Note that such a situation is not a common one and gives rise to two quite peculiar effects.
The first one is that along the real axis the resonance signal is purely real because
$g^2/(s-s_R)+{g^*}^2/(s-s_R^*)=2 {\Re g^2/(s-s_R)}$ (with $s$ the usual Mandelstam variable).
The second effect is that if $s_{{\rm th}}$ is the threshold represented in the figure then
  the maximum at the cusp is given by the sum of the two poles contributions
  $g^2/(s_{{\rm th}}-s_R)+{g^*}^2/(s_{{\rm th}}-s_R^*)$.
Therefore, if $s_R\to s_{{\rm th}}$ the
  cusp is enhanced compared to standard cases with only one pole for the virtual state.
 This situation could be well that corresponding to the $X(3872)$, as shown in Ref.~\cite{kang.170930.1},
 for which case the state whose threshold is depicted in Fig.~\ref{fig.171124.1}
 by the crossing of the two perpendicular 
 lines is the $\bar{D}^0{D^*}^0$. Furthermore, the $X(3872)$ has a tiny decay width to this state.
 As analyzed in Ref.~\cite{kang.170930.1}   the most recent data on spectral lines of the $X(3872)$
 are compatible with the resonance being a double- and even a triple-pole virtual state (in the limit of
 vanishing ${D^*}^0$ width). Thus, this resonance might be the first one 
 in particle phenomenology that manifests as a higher degree pole.
  In all the cases the poles emerge from the interplay between
  the presence of a bare state and the direct interactions between the $\bar{D}^0{D^*}^0$ mesons.
     Reference \cite{kang.170930.1} proceeds by making use of general principles of $S$-matrix theory. 
  Instead of considering explicitly the inclusion of a bare pole it derives the consequences of the presence
  of a near-threshold CDD pole, as an extra source   for a strong energy signal in the partial wave
  beyond the presence of the threshold.
  This is a more general approach than including a bare pole \cite{baru.171125.1}
  because it can also reproduce positive values for the effective range (as well as for
  other shape parameters in the effective range expansion) \cite{kang.170930.1}.


\end{document}